# Real-Space Adaptive-Coordinate Electronic Structure Calculations

François Gygi and Giulia Galli
*Institut Romand de Recherche Numérique en Physique des Matériaux (IRRMA)*
*CH-1015 Lausanne, Switzerland*

We present a real-space adaptive-coordinate method, which combines the advantages of the finite-difference approach with the accuracy and flexibility of the adaptive coordinate method. The discretized Kohn-Sham equations are written in generalized curvilinear coordinates and solved self-consistently by means of an iterative approach. The Poisson equation is solved in real space using the Multigrid algorithm. We implemented the method on a massively parallel computer, and applied it to the calculation of the equilibrium geometry and harmonic vibrational frequencies of the $CO_2$, CO, $N_2$ and $F_2$ molecules, yielding excellent agreement with the results of accurate quantum chemistry and Local Density Functional calculations.

Recently, important efforts in the development of pseudopotential ab initio electronic structure methods have focused on the ability to treat all elements of the periodic table on an equal footing, and to perform efficiently molecular dynamics (MD) simulations. In particular, the adaptive coordinate method[1], proposed to improve the efficiency of the widely used plane wave approach, has been shown to allow for accurate MD simulations including first-row elements[2]. The applicability of this method to the calculation of structural properties of solids[3] and to the computation of band structures[4] has also been demonstrated.

Another recent development in electronic structure calculations, the finite-difference (FD) real-space method[5], has tackled a serious drawback of the plane wave approach, i.e. its inability to describe non-periodic systems such as clusters or molecules. This problem has been usually circumvented by using very large unit cells in order to minimize the interaction between replicas of the system. However, this often leads to very slow convergence of the results as the cell size is increased, and can be forbidding for charged systems. The FD real-space method is based on a discretization of the Kohn-Sham equations, and does not imply any assumption of periodicity of the solutions, thus allowing one to choose boundary conditions (periodic or Dirichlet) most appropriate to the system under study. It was pointed out in Ref. 5 that the finite-difference approach might be well adapted to modern parallel computer architectures since it avoids completely the use of Fourier transforms. We note that an important feature of real-space methods is that they can lead to a straightforward implementation of linear scaling electronic structure algorithms (O(N) methods) within a self-consistent Density Functional approach[6]. Such algorithms can be naturally formulated in real space, since they make use of localized electronic orbitals defined in domains smaller than the entire simulation cell.

For the treatment of first row elements, the finite-difference approach suffers from the same inefficiency problem as plane waves. Indeed, in order to systematically improve the accuracy of a calculation, the grid spacing $h$ must be reduced on the entire grid, even though an increased resolution would be needed only in regions of rapidly varying potential. In the case of first-row elements this can be computationally very demanding. However, a *local* reduction of the grid spacing can be achieved by applying the concept of adaptive curvilinear coordinates to the finite-difference approach.

In this paper, we present a real-space adaptive-coordinate method, which combines the advantages of the FD approach with the accuracy and flexibility of adaptive coordinates. The discretized Kohn-Sham equations are rewritten in generalized curvilinear coordinates and solved self-consistently using an iterative approach. The method is implemented on a massively parallel computer, and applied to the calculation of the equilibrium geometry and harmonic vibrational frequencies of the $CO_2$, CO, $N_2$ and $F_2$ molecules, yielding excellent agreement with the results of other accurate first-principles calculations. Our results show that the use of curvilinear coordinates is essential to obtain a sufficiently small grid spacing in the vicinity of atoms, while keeping the numerical effort limited.

Curvilinear coordinates can be chosen to be either fully adaptive[1,3,4], or defined from given deformation functions as in Ref. 2, which is more efficient in molecular dynamics simulations. We follow Ref. 2, and define a coordinate transformation $\xi \to \mathbf{x}(\xi)$ by

$$\xi^i = x^i + \sum_\alpha (x^i - R_\alpha^i) f_\alpha(|\mathbf{x} - \mathbf{R}_\alpha|) \qquad (1)$$

where $\mathbf{R}_\alpha$ denote the atomic positions and the deformation functions $f_\alpha(r)$ are given by

$$f_\alpha(r) = A_\alpha \frac{a_\alpha}{r} \tanh \frac{r}{a_\alpha} \exp\left[-\left(\frac{r}{b_\alpha}\right)^2\right] \qquad (2)$$

This coordinate transformation maps a regular grid in $\xi$-space onto a curvilinear grid in (real) $\mathbf{x}$-space, and reduces the grid spacing in $\mathbf{x}$-space near atoms. In Eq. 2 $a_\alpha$ is the range of enhancement of the resolution around atom $\alpha$, and $b_\alpha$ determines a distance over which Euclidean coordinates are recovered. The local reduction in grid spacing at the center of an atom of species $\alpha$ is given by $h_{\text{eff}}/h_0 = 1/(1+A_\alpha)$, where $h_0$ denotes the grid spacing in Euclidean coordinates. The corresponding local effective energy cutoff at the atomic site is

$$E_{\text{cut}}^{\text{eff}} = \frac{\hbar^2}{2m}\left(\frac{\pi}{h_{\text{eff}}}\right)^2. \quad (3)$$

The wavefunctions are then written as the product[7]

$$\psi(\mathbf{x}) = \phi(\mathbf{x})\left|\frac{\partial\xi}{\partial x}\right|^{\frac{1}{2}} \quad (4)$$

where $|\partial\xi/\partial x|$ is the Jacobian of the transformation $\mathbf{x} \to \xi$. The function $\phi$, which was expanded in Fourier series in the plane-wave adaptive coordinate method[1] is now represented by its values on the curvilinear grid. The product representation Eq. (4) of $\psi$ has the advantage of conserving scalar products during a change of coordinate transformation $\mathbf{x}(\xi) \to \mathbf{x}'(\xi)$. This has the important consequence that a set of orthonormal wavefunctions remains orthonormal when $\mathbf{x}(\xi)$ changes due to the motion of atoms in a molecular dynamics simulation.

We write the curvilinear Laplacian as

$$\Delta = g^{ij}\frac{\partial}{\partial\xi^i}\frac{\partial}{\partial\xi^j} + \frac{\partial^2\xi^i}{\partial x^k \partial x^k}\frac{\partial}{\partial\xi^i} \quad (5)$$

where we expand the second term on the r.h.s. as

$$\frac{\partial^2\xi^i}{\partial x^k \partial x^k} = \frac{\partial\xi^l}{\partial x^k}\frac{\partial}{\partial\xi^l}\left(\frac{\partial\xi^i}{\partial x^k}\right). \quad (6)$$

Summation over repeated indices is implied throughout this paper. The discretized version of $\Delta$ is obtained by replacing all derivatives with respect to $\xi^i$ in (5) and (6) by finite-difference formulae and by using the analytic expression of $\partial\xi^i/\partial x^j$ derived from the definition of $\xi(\mathbf{x})$. In Eq.(5), $g^{ij}$ is the inverse of the Riemann metric tensor and is obtained directly from the analytic expression of $\partial\xi^i/\partial x^j$. Note that in Euclidean coordinates, the discretized Laplacian is self-adjoint and positive-definite to all orders in $h$. On the contrary, the discretized *curvilinear* Laplacian is only self-adjoint and positive-definite up to terms of the order of the truncation error (i.e. $O(h^N)$, where $N$ is the order of the finite-difference formula). In practical applications, this departure from self-adjointness and positive-definiteness has no noticeable effect on the convergence of the results, if the deformation functions $f_\alpha$ are sufficiently smooth. The truncation error can however become significant if the Jacobian $\partial\xi^i/\partial x^j$ is varying too rapidly on the scale of the grid spacing. We use finite-difference formulae of order 2 and 4 for the discretized derivatives i.e. 3- and 5-point formulae in each direction. Higher order finite difference formulae are not needed since the reduction in grid spacing provided by the adaptive coordinates usually brings sufficient accuracy. Low-order formulae also reduce the amount of interprocessor communications on a parallel computer.

The potential energy operator consists of norm-conserving, non-local pseudopotentials in the Kleinman-Bylander[8] form. The integrals associated with the local potential, non-local potential, and exchange correlation energy are evaluated by straightforward sums over the grid points, appropriately weighted with the Jacobian.

The properties of the curvlinear discretized Hamiltonian differ in many respects from those of the plane-wave Hamiltonian. First, the wavefunctions, the charge density and the potentials are all represented on the same grid, whereas in the plane-wave method, the potentials and the wave-functions are usually represented as Fourier series having different energy cutoffs. Furthermore, finite-difference methods do not rely on an expansion of the solutions on a basis set, and the energy therefore does not obey a variational principle. Instead, the truncation error of the energy can have an arbitrary sign, and in practice, the convergence of the energy to the exact value as $h \to 0$ is often attained from *below*. Finally, unlike in the plane-wave approach, the discretized Hamiltonian is not in general invariant under uniform translations of the system with respect to the position of the grid. This loss of translational invariance requires special care in the convergence of the calculation if spurious forces are to be avoided. In Euclidean coordinates, it is easily shown[9] that translational invariance of the energy is restored if the ionic pseudopotential is bandwidth limited, i.e. if its Fourier transform $V(q)$ vanishes beyond the Nyquist critical wave vector $q_c = \pi/h$. In practice, since typical norm-conserving pseudopotentials are not bandwidth limited, translational invariance of the energy can be recovered by either reducing $h$ until $V(q_c)$ becomes negligible, or by appropriately filtering the potential. Although no simple equivalent to the Nyquist critical wave vector $q_c$ exists in curvilinear coordinates, we found that a *local* reduction of the grid spacing $h$ near atoms, i.e. where the ionic potential has rapid oscillations, can be sufficient to restore the translational invariance of the energy.

Within the adaptive coordinate approach, the Poisson equation must be solved by an iterative method, since the curvilinear Laplacian is not diagonal in either real or reciprocal space. We solve Poisson's equation entirely in real space where the discretized curvilinear Laplacian is sparse. This allows one to work in domains of arbitrary shapes, with arbitrary boundary conditions. It also avoids the use of Fourier transforms, which is advantageous on massively parallel computers. As an iterative procedure, we choose the Multigrid method[10], which has the property of including an implicit preconditioning on all length scales, and is therefore rapidly convergent. In order to ensure the convergence of the Multigrid iteration, the discretized Laplacian operator must be non-singular. This can be accomplished by projecting appropriately the solution out of the nullspace of the discretized Laplacian. For our choice of the discretized Laplacian Eq.(5), and for periodic boundary conditions, the nullspace reduces to the constant function.

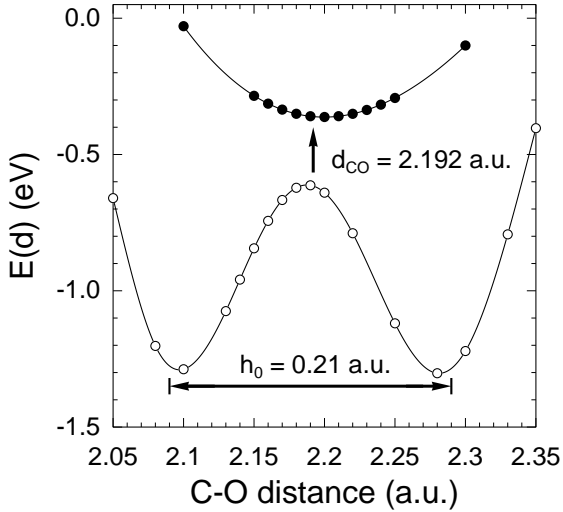

FIG. 1. Total energy of the $CO_2$ molecule as a function of the C-O distance in the linear conformation, calculated using Euclidean coordinates (open circles) and adaptive curvilinear coordinates (dots). Spline interpolations are shown as a guide to the eye. The two curves are shifted arbitrarily in order to allow for their comparison. The experimental value of the equilibrium C-O distance (arrow) and the grid spacing in Euclidean coordinates are also indicated.

The implementation of the real-space adaptive-coordinate approach was carried out on a Cray T3D massively parallel computer, using the Parallel Virtual Machine (PVM) environment. For simplicity, the simulation cell was divided into domains of equal shape and size which were distributed among processors. Other choices could be made to improve e.g. load balancing of the computations. Communications are limited to nearest neighbouring processors, and only imply the transfer of data defined on domain boundary layers. In particular, no global communication of a function defined on the entire grid is ever needed.

We applied the real-space adaptive-coordinate approach to the calculation of the equilibrium configuration and of the harmonic vibrational frequencies of the $CO_2$, CO, $N_2$ and $F_2$ molecules, within the local density approximation (LDA) of density functional theory. The experimental[11,12] equilibrium bond lengths in $CO_2$, CO and $N_2$ are 2.19, 2.13 and 2.07 a.u. respectively. Reproducing such short bond lengths, as well as vibrational frequencies in these molecules represents a stringent test for pseudopotentials. The description of $F_2$ is also a difficult test for a pseudopotential approach since the fluorine $p$ wavefunction is one of the most strongly localized valence wavefunction of all elements in the Periodic Table.

For these reasons, we used the norm-conserving pseudopotentials of Bachelet, Hamann and Schlüter[13], which reproduce accurately the atomic all-electron valence wavefunctions already at distances of 0.8, 0.7, 0.6 and 0.5 a.u. from the nuclei, for carbon, nitrogen, oxygen and fluorine respectively. We first computed the total energy of $CO_2$ as a function of the C-O distance $d_{CO}$ using Euclidean coordinates and a grid of $96 \times 72 \times 72$ points in a cell of $20 \times 15 \times 15$ a.u. This resolution would correspond to an effective energy cutoff of 227 Ry in the expansion of wavefunctions, charge density and potentials in a conventional plane wave calculation. We used a 5-point finite-difference formula in each direction to compute derivatives. The resulting total energies are shown on Fig. 1 (open circles), together with a spline interpolation passing through calculated points (solid line). The loss of translational invariance manifests itself strikingly by the presence of *two* minima in $E(d_{CO})$ in the vicinity of the experimental value of $d_{CO}$. The distance between the two minima is close to the grid spacing $h_0$, which confirms that the oscillations in $E(d_{CO})$ are spurious and depend on the relative positions of the atoms with respect to grid points. This clearly demonstrates that the resolution used in Euclidean coordinates is insufficient for a calculation of the equilibrium value of $d_{CO}$. We then repeated the calculation using the same number of grid points and the same unit cell, but with curvilinear coordinates, defined by the deformations given in Eq. (2) with $A_C = 0.258$, $a_C = 1.10$ a.u. and $b_C = 4$ a.u. for carbon, and $A_O = 0.989$, $a_O = 0.60$ a.u. and $b_O = 4$ a.u. for oxygen. This corresponds to effective energy cutoffs on the atomic sites of 360 Ry and 900 Ry for carbon and oxygen respectively. The total energies obtained with curvilinear coordinates are also shown in Fig. 1 (dots), together with a spline interpolation (solid line). The curve $E(d_{CO})$ has only one minimum at $d_{CO} = 2.198$ a.u. which compares nicely with the experimental value of the equilibrium bond length of 2.192 a.u. The harmonic vibrational frequency of the $\sigma_g^+$ symmetric stretching mode was extracted from a fourth order polynomial fit to $E(d_{CO})$. A similar calculation was carried out to determine the harmonic frequencies of the $\pi_u$ bending mode and of the $\sigma_u^+$ asymmetric stretching mode. We note that harmonic frequencies cannot be compared directly with experimental frequencies, due to the presence of anharmonic terms in the potential energy surface, especially for the $\sigma_g^+$ mode which is split by a Fermi resonance[14]. They can be compared, however, to harmonic frequencies obtained from a parametrization of the potential energy surface of $CO_2$ adjusted to experimental spectra[11,15]. Our calculated frequencies are 1336, 648 and 2374 cm$^{-1}$ for the $\sigma_g^+$, $\pi_u$ and $\sigma_u^+$ modes, respectively, which compare well with the corresponding values extracted from experiment (1354, 673 and 2397 cm$^{-1}$). A summary of the results for $CO_2$, and a comparison with other calculations are given in Table I.

We also computed the equilibrium bond lengths and the harmonic vibrational frequencies of the CO, $N_2$, and $F_2$ molecules. For $N_2$, and $F_2$, we used deformation functions $f_\alpha$ yielding an effective energy cutoff of 800 and 990 Ry at the atomic sites respectively.

TABLE I. Equilibrium C-O distance $d_{\rm CO}$ and harmonic vibrational frequencies in $CO_2$ calculated using the real-space adaptive-coordinate method, compared with harmonic frequencies adjusted to experimental spectra, and with the results of other calculations (LDA, Hartree-Fock (SCF), configuration interaction (CISD) and coupled cluster (CCSD,CCSD(T)). Frequencies are given in $cm^{-1}$, and distances in a.u.

|  | $d_{\rm CO}$ | $\pi_u$ | $\sigma_g^+$ | $\sigma_u^+$ |
| --- | --- | --- | --- | --- |
| LDA[a] | 2.220 | | | |
| SCF[b] | 2.145 | 779 | 1513 | 2556 |
| CISD[b] | 2.167 | 729 | 1442 | 2502 |
| CCSD[b] | 2.184 | 697 | 1389 | 2432 |
| CCSD(T)[b] | 2.198 | 672 | 1345 | 2391 |
| This work | 2.198 | 648 | 1336 | 2374 |
| Expt.[b] | 2.192 | 673 | 1354 | 2397 |

[a]From Ref. 17.
[b]From Ref. 11.

Our results are summarized in Table II where they are compared with those of other LDA calculations[16,17] and with experiment. The overall agreement with other LDA results is excellent. The agreement with experiment is also remarkably good, except for $F_2$, for which a more accurate treatment of correlation is needed.

In conclusion, we have presented a real-space adaptive-coordinate method for electronic structure calculations. This approach combines the advantages of the finite-difference method, with the capability of the adaptive-coordinate method to treat first-row and heavier elements on the same footing. We have shown that the Poisson equation in curvilinear coordinates can be solved efficiently using the Multigrid method, and that a *local* reduction of the grid spacing near atoms can restore the translational invariance of the energy. The method was implemented on a massively parallel computer, and applied to the calculation of the equilibrium bond lengths and vibrational frequencies of several molecules containing first-row elements, yielding excellent agreement with other accurate first principles results. Work is in progress to apply the method to ab-initio molecular dynamics simulations, and to extend it to a linear scaling formulation.

We would like to thank A. Pasquarello and R. Resta for useful comments on the manuscript. We acknowledge support from the Swiss National Science Foundation under grant 20-39528.93. This work is partially supported through the Parallel Application Technology Program (PATP) between EPFL and Cray Research, Inc.

TABLE II. Equilibrium bond lengths $d$ and harmonic vibrational frequencies $\omega_e$ in CO, $N_2$ and $F_2$ calculated using the real-space adaptive-coordinate method, compared with other LDA results and with values extracted from experiment.

|  | CO | $N_2$ | $F_2$ |
| --- | --- | --- | --- |
| $d$ (a.u.) | | | |
| LDA | 2.162[b] | 2.08[a],2.11[b] | 2.62[a],2.63[b] |
| This work | 2.132 | 2.071 | 2.635 |
| Expt. | 2.132[c] | 2.07[a] | 2.68[a] |
| $\omega_e$ ($cm^{-1}$) | | | |
| LDA | | 2387[a] | 1075[a] |
| This work | 2151 | 2382 | 1051 |
| Expt. | 2170[c] | 2358[a] | 892[a] |

[a]From Ref. 16.
[b]From Ref. 17.
[c]From Ref. 12.